\documentclass{iau}
\usepackage{graphicx,natbib}
 
\newcommand{\apj}{ApJ}           
\newcommand{\apjl}{ApJ}           
\newcommand{\mnras}{MNRAS}       
\newcommand{\nat}{Nature}
\newcommand{\aap}{A\&A}
\newcommand{\araa}{ARA\&A}
\newcommand{\aj}{AJ}

\newcommand{\aapr}{A\&A Rev.}

\newcommand{\HI}{{\sc H\,i}}
\newcommand{\arcsec}{^{\prime\prime}}
\newcommand{\degr}{^\circ}

\title{The WSRT HALOGAS Survey}

\author[Heald et al.]{George Heald$^{1,2}$ \and the HALOGAS Team}

\affiliation{$^1$ASTRON, Postbus 2, 7990 AA Dwingeloo, The Netherlands\\
$^2$Kapteyn Astronomical Institute, Postbus 800, 9700 AV, Groningen, The Netherlands\\
email: {\tt heald@astron.nl} }

\pubyear{2014}
\volume{309}
\jname{Galaxies in 3D across the Universe}
\editors{B. L. Ziegler, F. Combes, H. Dannerbauer, M. Verdugo, eds.}

\begin{document}

\maketitle

\begin{abstract}
We present an overview of the HALOGAS (Hydrogen Accretion in LOcal GAlaxieS) Survey, which is the deepest systematic investigation of cold gas accretion in nearby spiral galaxies to date. Using the deep \HI\ data that form the core of the survey, we are able to detect neutral hydrogen down to a typical column density limit of about $10^{19}\,\mathrm{cm^{-2}}$ and thereby characterize the low surface brightness extra-planar and anomalous-velocity neutral gas in nearby galaxies with excellent spatial and velocity resolution. Through comparison with sophisticated kinematic modeling, our 3D HALOGAS data also allow us to investigate the disk structure and dynamics in unprecedented detail for a sample of this size. Key scientific results from HALOGAS include new insight into the connection between the star formation properties of galaxies and their extended gaseous media, while the developing HALOGAS catalogue of cold gas clouds and streams provides important insight into the accretion history of nearby spirals. We conclude by motivating some of the unresolved questions to be addressed using forthcoming 3D surveys with the modern generation of radio telescopes.
\keywords{galaxies: spiral --- galaxies: evolution --- galaxies: kinematics and dynamics}
\end{abstract}

\firstsection
\section{Background}

Spiral galaxies require a continuous supply of gas to maintain star formation over long timescales and to explain their metallicity distribution and evolution \citep[see][and references therein]{sancisi_etal_2008}. What form this gas supply takes has not been observationally established, but it may be that the dominant form of supply is cold gas accretion \citep[e.g.,][]{keres_etal_2009}. In the Milky Way (MW) and the nearest galaxies high velocity clouds (HVCs) are observed \citep{wakker_vanwoerden_1997,thilker_etal_2004}, some with low metallicity \citep{wakker_etal_1999,collins_etal_2007}. The HVCs may contribute a small fraction of the fuel needed for star formation, but a more complete census is required to be able to draw general conclusions. While distinct \HI\ clouds of this nature are not commonly detected in the outskirts of galaxies \citep[e.g.,][]{giovanelli_etal_2007,putman_etal_2012}, another morphological feature -- thick gas disks -- is becoming increasingly recognized as a typical structural feature of spiral galaxies \citep[e.g.,][]{oosterloo_etal_2007,putman_etal_2012}. These gas disks can contribute up to $\approx10-20\%$ of the total \HI\ mass of their host galaxies, and have distinct kinematics \citep[e.g.,][]{fraternali_etal_2002}. These slowly rotating thick disks may be connected to the accretion history \citep{marinacci_etal_2010} and provide an observational tracer of coronal material feeding the disk.

Searching for clouds and for thick \HI\ disks requires specialized deep observations. The HALOGAS \citep[Hydrogen Accretion in Local Galaxies;][]{heald_etal_2011} Survey was undertaken to provide the required capability for a large enough sample to begin to draw statistical conclusions. The core of the survey are Westerbork Synthesis Radio Telescope (WSRT) observations of 22 edge-on and moderately-inclined nearby galaxies, each observed for 120 hours in the 21-cm \HI\ line. The typical $5\sigma$ column density sensitivity is $N_\mathrm{HI}\lesssim10^{19}\,\mathrm{cm^{-2}}$ for typical linewidth of $\Delta\,v=12\,\mathrm{km\,s^{-1}}$, making HALOGAS the deepest interferometric galaxy \HI\ survey available to date. HALOGAS has an optimal angular resolution of $\approx30^{\prime\prime}$ (1.5 kpc at a typical distance of 10 Mpc), unsuitable for detailed high-resolution studies of the gas-star formation connection like THINGS \citep[The \HI\ Nearby Galaxy Survey;][]{walter_etal_2008}, but ideal for faint diffuse emission. The typical HALOGAS mass sensitivity for unresolved clouds with the same linewidth allows the detection of HVC analogues in the outskirts of the survey galaxies,
\begin{equation}
M_\mathrm{HI}=2.7\times10^5\,\left(\frac{D}{10\,\mathrm{Mpc}}\right)^2\,M_\odot.
\end{equation}

The primary HALOGAS observations were completed as of early 2013. A number of ancillary data products have been collected to supplement our deep \HI\ observations. Deep multi-band optical imagery have been obtained at the {\it Isaac Newton Telescope} (INT) and Kitt Peak National Observatory (KPNO); see e.g. \citet{deblok_etal_2014}. Sensitive {\it GALEX} UV data have also been collected to supplement deep survey data already available for some HALOGAS targets. Together with the \HI\ line observations, broadband full-polarization continuum data are also available for many of the survey galaxies, allowing the detailed investigation of magnetic fields and cosmic rays in the same galaxies. The interferometric HALOGAS data are being supplemented with single dish data from Effelsberg and the the Green Bank Telescope (GBT) --- useful not only to provide a missing short-spacing correction, but to enhance the search for clouds and streams in the outer parts of HALOGAS galaxies.

\section{Overview of HALOGAS results}\label{section:overview}


Sophisticated data analysis techniques are required in order to achieve reliable structural parameters from the sensitive \HI\ data cubes provided by HALOGAS. Isolating thick \HI\ disks in particular requires detailed modeling of the 3D gas distribution and kinematics. This modeling work is mostly performed using the TiRiFiC \citep[Tilted Ring Fitting Code;][]{jozsa_etal_2007} software. 

\begin{figure}
\centering
\includegraphics[width=0.49\hsize]{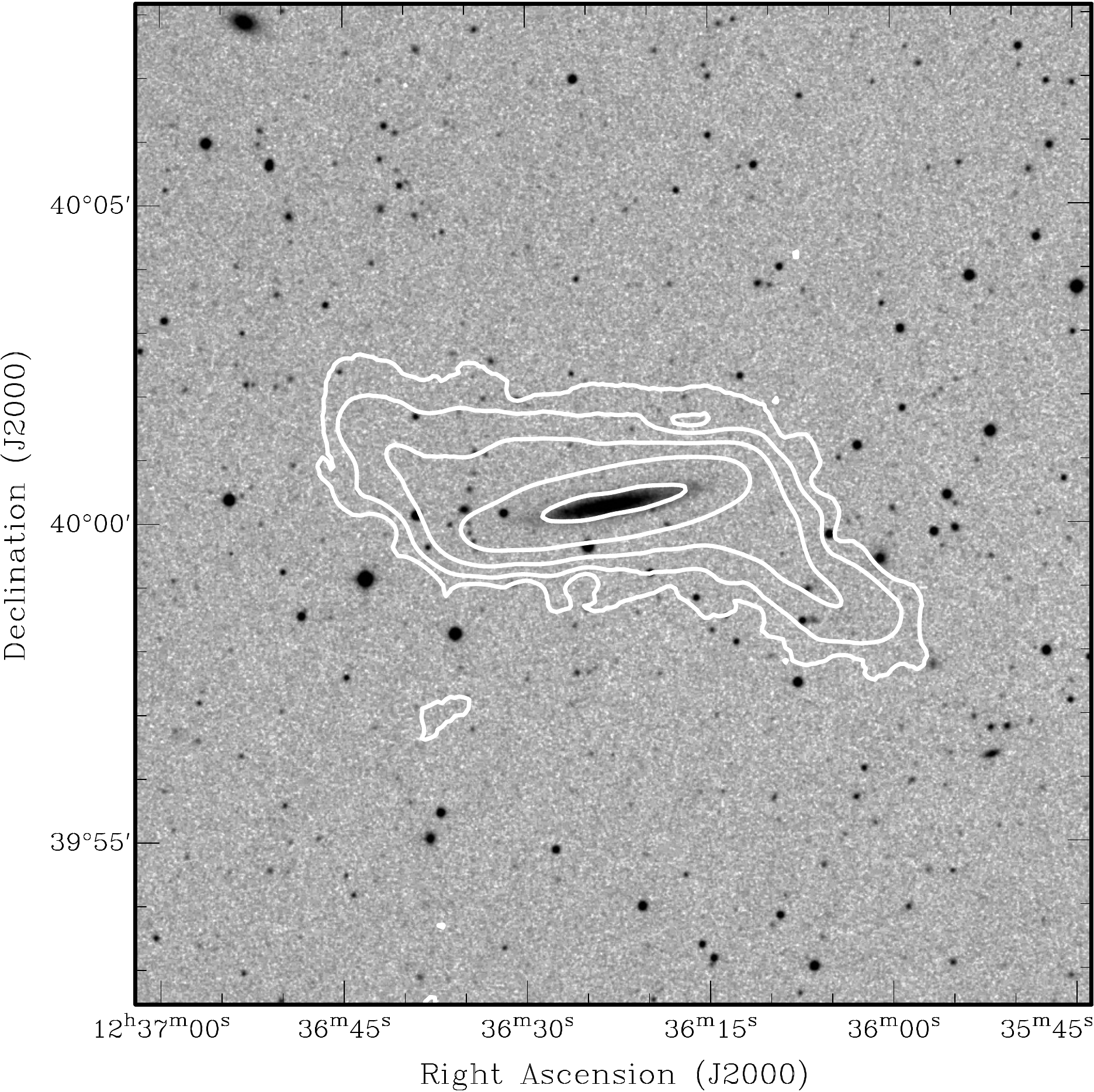} \hfill \includegraphics[width=0.49\hsize]{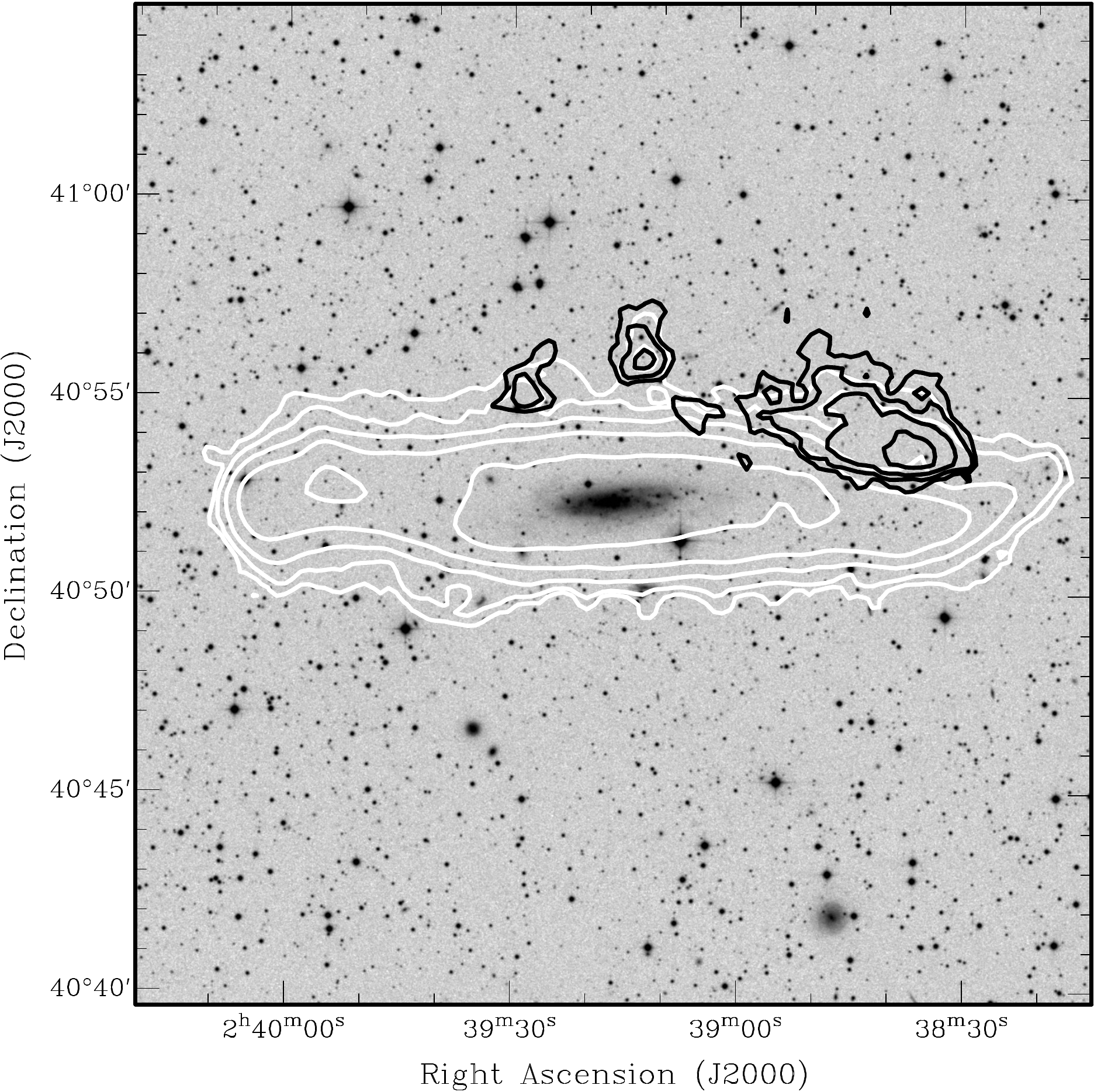}
\caption{Two HALOGAS sample galaxies in optical and \HI. {\it Left}: DSS2 $R$-band image of UGC~7774 displayed on a logarithmic stretch and overlaid with HALOGAS \HI\ total intensity contours, starting at $N_\mathrm{HI}=9\times10^{18}\,\mathrm{cm^{-2}}$ and increasing by multiples of 2. The angular resolution of the \HI\ data is $36\arcsec\times33\arcsec$. {\it Right}: DSS2 $R$-band image of NGC~1003 displayed on a logarithmic stretch and overlaid with HALOGAS \HI\ total intensity contours. The \HI\ from the entire galaxy is shown in white contours, starting at $N_\mathrm{HI}=5\times10^{18}\,\mathrm{cm^{-2}}$ and increasing by multiples of 4. The black contours show the HVC analogues discussed in the text (\S\,\ref{subsection:accretion}), starting at about the same column density and increasing by multiples of 2. The resolution of the \HI\ data is $39\arcsec\times34\arcsec$.}\label{fig:fig1}
\end{figure}

Two examples that illustrate the need for detailed tilted ring modeling are shown in Figure \ref{fig:fig1}. In the case of UGC~7774, the strong warping already apparent in the plane of the sky \citep[see also][]{garciaruiz_etal_2002} also causes the vertical thickening. In the other galaxy, NGC~1003, the disk structure recovered from a detailed 3D analysis shows a substantial warp ($\approx15-20\degr$) along the line of sight, causing the disk to appear thicker in projection than the actual vertical distribution of the \HI\ layer.

Another interesting example, NGC~5023 (not shown), illustrates an extreme case: recovery of spiral structure in an edge-on disk \citep[see][]{kamphuis_etal_2013}. The modeling works best when these spiral structures are present and are required to fully understand the kinematics, emphasizing that this level of detail is essential to properly describe the 3D structure of the \HI\ in galaxies. 

HALOGAS galaxies are both found to host thick \HI\ disks \citep{gentile_etal_2013,deblok_etal_2014} and to have little extraplanar \HI\ \citep{zschaechner_etal_2011,zschaechner_etal_2012}. This dichotomy provides powerful leverage on the origin of \HI\ thick disks in galaxies.

\subsection{Star formation and the origin of thick \HI\ disks}

The entire HALOGAS sample has been inspected in order to characterize the presence and characteristics of extraplanar \HI. We use both the 3D modeling mentioned in \S\,\ref{section:overview}, as well as an approximate disk-halo separation of the type presented by \citet{fraternali_etal_2002}. On this basis we have searched for evidence of a common origin of the extraplanar \HI\ by comparing with the general properties of the sample galaxies. From the HALOGAS sample we find a tentative connection between the star formation rate density and the presence of thick \HI\ disks, such that thick \HI\ disks appear to be present in galaxies above a threshold SF energy injection. We note that a similar relationship has been identified for radio continuum halos \citep{dahlem_etal_2006}. In the case of the \HI, there appears to be a strong dependence on the star formation rate density, but a weaker dependence on the stellar mass density that traces the strength of the gravitational potential. 

\subsection{Cold gas accretion in galaxies}\label{subsection:accretion}

Another key result of the HALOGAS Survey is the identification of \HI\ clouds and streams in the vicinity of the sample galaxies that could originate through the cold gas accretion process, and to determine the impact on the star formation history of the sample. As previously noted, the occurrence of gas clouds in the outskirts of galaxies is low. The HALOGAS Survey, despite its high sensitivity, does not substantially change this picture. The HALOGAS catalog of isolated gas features is still in preparation (as we work toward a statistically sound census across the full sample), but we can already state qualitatively that there is a low incidence of clouds and streams. The contribution of visible \HI\ accretion to the star formation fueling process in nearby galaxies appears to be minimal.

As a representative example, we highlight NGC~1003 as shown in Figure~\ref{fig:fig1}. A small number of morphologically and kinematically distinct gas features have been identified from an inspection of the 3D \HI\ dataset; these are shown with black contours in Fig.~\ref{fig:fig1}. These features do not appear to have stellar counterparts, even in our deep optical images. The \HI\ masses and distances from NGC~1003 of these clouds are similar to the properties of the HVCs around the MW. The mass of these HVC analogues adds up to only $M_\mathrm{HI}=4\times10^6\,M_\odot$, which over a dynamical time ($\tau_\mathrm{dyn}\approx500\,\mathrm{Myr}$) contributes only 2\% of the star formation rate of NGC~1003 \citep[$\mathrm{SFR}=0.40\,M_\odot\,\mathrm{yr^{-1}}$;][]{heald_etal_2012}. Alternatively, if the gas contained in these clouds were used to completely fuel the current SFR, they would need to be replenished after only $\tau_\mathrm{repl}\approx10\,\mathrm{Myr}$. These rates neglect a possible contribution from an ionized gas component to which we do not have observational constraints. We note however that the contribution of ionized gas to the total mass of MW HVCs can be substantial \citep[e.g.,][]{lehner_howk_2011,fox_etal_2014}; QSO absorption line studies of low column density \HI\ probed by HALOGAS where possible would be very useful to constrain this aspect of the mass budget.

\section{Future prospects}

The scientific themes addressed by the HALOGAS project will help to focus the groundbreaking science questions to be addressed with the Square Kilometre Array (SKA) and its pathfinder and precursor projects. For descriptions of surveys targeting deeper observations of larger nearby galaxy samples, as well as investigations of low column density IGM material, see the discussions by \citet{deblok_etal_2015} and \citet{popping_etal_2015}, respectively.

\section*{Acknowledgements}

\noindent
The Westerbork Synthesis Radio Telescope is operated by ASTRON (Netherlands Institute for Radio Astronomy) with support from the Netherlands Foundation for Scientific Research (NWO).

\end{document}